\documentclass[12pt]{iopart}

\makeatletter
\@namedef{ver@amsmath.sty}{}
\makeatother
\usepackage{amstext}
\usepackage[version=3]{mhchem}
\usepackage{graphicx}
\usepackage{epstopdf} 
\usepackage{amssymb}
\newcommand{\ket}[2][]{{|#2\rangle_{#1}}}
\newcommand{\bra}[2][]{{}_{#1}\langle #2|}
\def\ben{\begin{eqnarray}}
\def\een{\end{eqnarray}}
\def\bei{\begin{itemize}}
\def\eei{\end{itemize}}

\def\<{\langle}
\def\>{\rangle}

\def\cl{\frac{\cos(\omega_Lt)}{\omega_L^2-\Omega^2}}
\def\cu{\frac{\cos(\omega_Ut)}{\omega_U^2-\Omega^2}}
\def\sl{\frac{\sin(\omega_Lt)}{\omega_L^2-\Omega^2}}
\def\su{\frac{\sin(\omega_Ut)}{\omega_U^2-\Omega^2}}
\def\com{\cos(\Omega t)}
\def\som{\sin(\Omega t)}
\def\ul{\frac{\omega_L}{\omega_L^2-\Omega^2}}
\def\uu{\frac{\omega_U}{\omega_U^2-\Omega^2}}
\def\loga{\log\frac{(\omega_L-\Omega)(\omega_U+\Omega)}{(\omega_L+\Omega)(\omega_U-\Omega)}}
\begin{document}

\newtheorem{thm}{Theorem}
\newtheorem{prop}[thm]{Proposition}
\newtheorem{cor}[thm]{Corollary}
\newtheorem{lem}[thm]{Lemma}
\newtheorem{fact}[thm]{Fact}
\newtheorem{remark}[thm]{Remark}

\title{Analytical studies of Spectrum Broadcast Structures in Quantum Brownian Motion}

\author{J.~Tuziemski$^{1,2}$, J.~K.~Korbicz $^{1,2}$}
\address{$^1$Faculty of Applied Physics and Mathematics, Gda\'nsk University of Technology, 80-233 Gda\'nsk, Poland}
\address{$^2$National Quantum Information Centre in Gda\'nsk, 81-824 Sopot, Poland}

\begin{abstract}
{Spectrum Broadcast Structures are a new and fresh concept in the quantum-to-classical transition, introduced recently in the context of 
decoherence and the appearance of objective features in quantum mechanics. These are specific quantum state structures, responsible
for an apparent objectivity of a decohered state of a system. Recently they have been shown to appear in the well known 
Quantum Brownian Motion model, however the final analysis relied on numerics. Here, after a presentation of the main concepts,  
we perform analytical studies of the model, showing the timescales and the efficiency 
of the spectrum broadcast structure formation. We consider a massive central system and a 
somewhat simplified environment being random with a 
uniform distribution of the frequencies. 
}
\end{abstract}

\maketitle
\section{Introduction}
A measurement in quantum mechanics typically alters a state of the system, so that if several observers try to
measure a certain observable they will in general interfere with each other. This is in stark contrast with classical world,
where properties of systems such as position, momentum, etc can be in principle observed by as many observers as one wishes, they will all
agree on the results (modulo eventual reference frame transformations), and moreover will not disturb the system. 
This observer-independence and non-disturbance may be taken as a basis of an intuitive definition of objectivity.
Thus a problem arises: How can one explain the observed objectivity of everyday world with quantum mechanics?
It can be seen as one of the aspects of the quantum-to-classical transition---a problem present from the very
beginning of quantum mechanics \cite{Schilpp}. 

One of the prominent attempts to address this problem has been known as quantum Darwinism \cite{ZurekNature}. It is a refined and more realistic version 
of decoherence theory  (see e.g. \cite{Schlosshauer,decoh,Petruccione}), where one realizes that often observations are 
made indirectly, through portions of the environment rather than by a direct interaction with the object (e.g. an illuminated object scatters photons, which are then detected by
the eyes of the observers). Hence, the environment is no longer treated merely as a source of noise and dissipation but is recognized as
an important "information carrier". This implies a paradigmatic shift in the main object of studies in the theory of open quantum systems---from the reduced state of the system 
$\varrho_S$ alone \cite{Schlosshauer,decoh,Petruccione} to a joint state of the system and an observed fraction $fE$ of the environment $\varrho_{S:fE}$.
Objectivity is then linked to "information redundancy":
If the environment acquires in the course of the decoherence a large number of copies of the state of the system 
and this information can be read out without disturbance, then the state of the system becomes objective \cite{ZurekNature}.
In pictorial terms, objective becomes information about the system which not only "survives" the interaction with the environment,
but manages to proliferate in the latter. 
As a measure of this effect a family of scalar criteria based on the quantum mutual information has been proposed.
The criteria check for a change in the quantum mutual information $I(\varrho_{S:fE})$ between the system and a fraction of the environment 
as a function of this fraction size (e.g. a number of scattered photons taken into account in $fE$). If at some point no change is produced while increasing the fraction size
(and the so called classical plateau appears on partial information plots), it is then concluded that environment stores a largely redundant 
amount of information about the system state (more precisely about pointer states to which the system decoheres).

This approach has been reconsidered in \cite{my} on a ground that scalar criteria, and quantum mutual information in particular,
might not be the right tools and more convincing arguments for information proliferation are needed, preferably on the most fundamental level available---that of quantum states.
Such an approach has indeed been proposed in \cite{my} (for a somewhat complementary approach see \cite{Pawel}). 
Starting from an intuitive definition of objectively existing state of the system \cite{ZurekNature, OPZ2004}
it has been shown that, under certain assumptions, it  singles out a structure of a partially reduced state of the system and the observed
fraction of the environment compatible with objectivity to the following one, called Spectrum Broadcast Structure (SBS):
\ben\label{sbs}
\varrho_{S:fE}=\sum_ip_i\ket i\bra i \otimes \varrho_i^{E_1}\otimes\dots\otimes\varrho_i^{E_M}, \quad \varrho_i^{E_k}\varrho_{i'\ne i}^{E_k}=0,
\een
where $\ket i$ is the pointer basis to which the system decoheres, $p_i$ are the initial pointer probabilities and 
$\varrho_i^{E_k}$ are some states of the fragments of the environment $E_1,\dots,E_M\in fE$, which are supposed to be observed
and hence cannot be traced out. The information about the state of the system--- the index $i$, is encoded in a number of copies $1,\dots,M$
in the environment through the states $\varrho_i^{E_k}$ and can be perfectly recovered due to the assumed non-overlap condition in
(\ref{sbs}) and without any disturbance (on average) to the whole state $\varrho_{S:fE}$. 
By the foregoing discussion this leads to an apparent objectivity 
of the state of the system. Quite surprisingly the converse is also true as shown in \cite{my} with a help of several assumptions.
One of them is that "non-disturbance" requirement 
should be understood in the sense of Bohr's reply \cite{Bohr} to the famous EPR paper \cite{EPR}
and which has been further formalized in \cite{Wiseman}. The other important assumption is  a, so called, strong independence condition, demanding that the only
correlation between parts of the environment is due to the common information about the system. 
Formally, from quantum information point of view,
states (\ref{sbs}) realize a certain weak form of state broadcasting \cite{broadcasting}, called spectrum broadcasting \cite{my_kanaly}: The spectrum $p_i$
of the reduced state of the system $\varrho_S$ is present in many copies in the environments $E_1,\dots,E_M$ and can be retrieved from there by projections on the
supports of $\varrho_i^{E_k}$ (due to the non-overlap condition in (\ref{sbs})). This broadcasting process can be also described by a channel \cite{my_kanaly}:
\ben\label{L}
\Lambda^{S\to fE}(\varrho_{0S})=\sum_i\langle i |\varrho_{S0}| i\rangle \varrho^{E_1}_i\otimes\cdots\otimes \varrho^{E_{M}}_i,
\een
where $\varrho_{0S}$ is the initial state of the system. Thus, objectivity can be seen a s a result of a certain broadcasting process given by the channel (\ref{L}).

The power of the above result is that it links objectivity and quantum state structures in a completely abstract, model independent way. A natural question then arises
if those structures appear in the canonical models of decoherence \cite{Schlosshauer}: collisional decoherence, Quantum Brownian Motion (boson-boson), spin-spin and spin-boson models. 
So far the first two models were analyzed  and the answer is in general affirmative: There exists parameter regimes of the models such that SBS are formed. 
The first studied example was the famous illuminated sphere model
of collisional decoherence of Joos and Zeh \cite{JoosZehsfera}. Following the quantum Darwinism inspired analysis of \cite{Zureksfera},
it has been shown in \cite{sfera} that indeed spectrum broadcast structures (\ref{sbs}) are asymptotically formed in  the course of the evolution, even if the 
environment is noisy (initially in a mixed state) and the appropriate time scales were given. Important, from the perspective of the current work, are the methods
introduced for checking for the SBS, which will be reviewed in the next chapter. The sphere model, however illustrative, is rather simple since the system has no self-dynamics.
More realistic, and richer, in this sense  is  Quantum Brownian Motion, where a central harmonic oscillator is linearly coupled to a bath of harmonic oscillators.
This is arguably one of the most popular models describing quantum dissipative systems.
Despite its long history \cite{Ullersma}, only recently studies of the informational content of the environment has appeared \cite{ZurekQBM,RoncagliaPaz,EPL,Photonics}.
The first two works analyzed both numerically and analytically (in the massive central system regime) the scalar condition of quantum Darwinism, assuming initially pure environment and
squeezed state of the system and showing that indeed the characteristic classical plateau is being formed. On the other hand in \cite{EPL,Photonics} the model has been analyzed from 
the SBS perspective and,  under somewhat similar conditions as above but with  thermal environment, a numerical evidence have been found that indeed there are
parameter regimes so that a SBS is formed. A distinctive feature of the found structure is that it is dynamical and evolves in time: The pointer basis in (\ref{sbs})
rotates in time according to the self-Hamiltonian of the central oscillator and at any instant a SBS is being formed, encoding traces of this motion.
One has to stress that due to the mentioned paradigmatic shift in the treatment of the environment, i.e. it  may contain useful information, one cannot assume it to be so inert
as not to feel the presence of the system, as it is done in the usual Born-Markov approximation and master equation approach to open quantum systems
(see e.g. \cite{Schlosshauer} for an introduction and standard applications). Thus, in particular our study of Quantum Brownian Motion does not rely on the Born-Markov approximation and 
master equation methods, but rather on a direct state analysis (the details are presented in Section \ref{qbm}). 
The drawback of the previous studies \cite{EPL} is that the analysis of the SBS formation was at the end performed numerically.
Here, continuing the previous research, we overcome this difficulty and show analytically that there is a parameter regime of the model
such that a dynamical SBS is formed.  We give analytical expressions for the decoherence and the SBS formation time-scales
in both low- and high-temperature regimes.

\section{Checking for Spectrum Broadcast Structures}\label{check}
The method of detection of SBS developed so far \cite{sfera, Photonics} is rather direct and most naturally apply to the situation  
when the system-environment interaction is of the von Neumann measurement type:
\ben\label{vnH}
\hat H_{SE} =\hat X\otimes \sum_{k=1}^N \hat Y_k,
\een
where $\hat X$, $\hat Y_k$ are some observables of the system and the $k$-th environment respectively, assumed for simplicity
to have discrete spectra. Albeit of a special form, this class of Hamiltonians is of a fundamental importance both for the
decoherence \cite{Schlosshauer} and measurement \cite{vN} theories and thus worth investigating.
To illustrate the method, we will neglect here the self-Hamiltonians of the system and the environment (quantum measurement limit)
as one can then calculate everything explicitly.
The resulting unitary evolution given by (\ref{vnH}) is of a controlled unitary type, where the system controls the environments
through eigenvalues $\xi$ of $\hat X$:
\ben\label{CU}
\hat U(t)=\sum_\xi \ket \xi\bra \xi\otimes \hat U_1(\xi;t)\otimes\cdots\otimes\hat U_N(\xi;t) ,\quad  \hat U_k(\xi;t)\equiv e^{-i\xi t\hat Y_k/\hbar}.
\een
Assuming, as it is usually done, a fully product initial state $\varrho_{0S}\otimes\varrho_{01}\otimes\cdots\otimes\varrho_{0k}$, one immediately obtains
that after the tracing of some portion of the environment, denoted $(1-f)E$ and containing a fraction $fN$, $0<f<1$ subsytems, the state reads:
\begin{eqnarray}
&&\varrho_{S:fE}(t)=tr_{(1-f)E}\left[\hat U(t)\varrho_{0S}\otimes\bigotimes_{k=1}^N\varrho_{0k}\hat U(t)^\dagger\right]\label{SfE}\\
&&=\sum_\xi\langle \xi|\varrho_{0S}|\xi\rangle \ket \xi \bra \xi\otimes \bigotimes_{k=1}^{fN} \hat U_k(\xi;t)\varrho_{0k}\hat U_k(\xi;t)\label{diag}\\
&&+\sum_{\xi\ne \xi'} \Gamma_{\xi,\xi'}(t) \langle \xi|\varrho_{0S}|\xi'\rangle \ket \xi \bra {\xi'}\otimes \bigotimes_{k=1}^{fN} \hat U_k(\xi;t)\varrho_{0k}\hat U_k(\xi';t),
\label{off}
\end{eqnarray}
where:
\ben
\Gamma_{\xi,\xi'}(t)\equiv\prod_{k\in (1-f)E} tr\left[\hat U_k(\xi;t)\varrho_{0k}\hat U_k(\xi';t)\right]=\prod_{k\in (1-f)E} tr\left[\varrho_{0k}e^{-i(\xi-\xi') t\hat Y_k/\hbar}\right]
\een
is the usual decoherence factor between the states $\ket \xi$, $\ket {\xi'}$. A check for the SBS (\ref{sbs}) proceeds in two steps.

First of all the coherent part (\ref{off}), containing entanglement between the system and the environment, should vanish and this is
of course the usual decoherence process, controlled by $\Gamma_{\xi,\xi'}(t)$. If one is able to show $|\Gamma_{\xi,\xi'}(t)|=0$ with time
for all different pair of $\xi,\xi'$, one proves that decoherence has taken place and $\ket \xi$ becomes the pointer basis.

Second, we check if the information deposited in the environment during the decoherence can be perfectly read out, i.e.
if the system-dependent states of the environments:
\ben\label{Estates}
\varrho_{\xi k}(t)\equiv U_k(\xi;t)\varrho_{0k}U_k(\xi;t)
\een
have non-overlapping supports (cf. (\ref{sbs})): 
\ben\label{orthogonal}
\varrho_{\xi k}(t)\varrho_{\xi'k}(t)= 0,
\een 
and hence are perfectly one-shot distinguishable. Among different measures of distinguishability \cite{Fuchs}, the most suitable turns out to be the
generalized overlap  
\ben\label{go}
B(\varrho_1,\varrho_2)\equiv tr\sqrt{\sqrt{\varrho_1}\varrho_2\sqrt{\varrho_1}}. 
\een
This is due to the fact that most interesting 
are the situations when the interaction with each individual portion of the environment in (\ref{vnH}) is vanishingly small (see e.g. \cite{JoosZehsfera}).
Then, one cannot expect (\ref{orthogonal}) to hold at the level of single environments. Right to the contrary, since each of the unitaries 
$\hat U_k(\xi;t)$ weakly depends on the parameter $\xi$, the states $\varrho_{\xi k}(t) $ are almost identical for different $\xi$'s.
In other words, the information about $\xi$ is diluted in the environment. However, it can happen that by grouping subsystems of the observed part of environment $fE$ into larger fractions, 
called macrofractions $mac$ and introduced in \cite{sfera}, one can approach the perfect distinguishability (\ref{orthogonal}) at the level of
macrofraction states $\varrho_{\xi}^{mac}(t)\equiv\bigotimes_{k\in mac}\varrho_{\xi k}(t)$. Generalized overlap is well suited for such tests 
due to its factorization with the tensor product: 
\ben\label{Bmac}
B^{mac}_{\xi,\xi'}(t)\equiv B\left(\varrho_{\xi}^{mac}(t),\varrho_{\xi'}^{mac}(t)\right)=\prod_{k\in mac}B\left(\varrho_{\xi k}(t),\varrho_{\xi' k}(t)\right).
\een

Summarizing, if one is able to prove that for some time both functions vanish
\ben\label{indic}
|\Gamma_{\xi,\xi'}(t)|\approx 0,\ B^{mac}_{\xi,\xi'}(t)\approx 0,
\een
then from (\ref{diag},\ref{off}) this is equivalent to the formation of the spectrum broadcast structure (\ref{sbs}):
\ben
\varrho_{S:fE}\approx\sum_\xi p_\xi \ket\xi\bra \xi\otimes \varrho^{mac_1}_\xi\otimes\cdots\otimes \varrho^{mac_M}_\xi,
\een
with $\varrho^{mac_k}_\xi$ having orthogonal supports for different $\xi$'s and the convergence is in the trace norm.

The introduced above grouping into macrofractions (or equivalently coarse-graining of the observed environment) 
can be seen as a reflection of detection thresholds of real-life detectors, e.g. an eye. 
Since one is usually interested in a thermodynamic-type of a limit $N\to\infty$ it is important
that those fractions scale with $N$ (hence the name "macrofractions"). 

\section{Spectrum Broadcast Structures in Quantum Brownian Motion}\label{qbm}
The model Hamiltonian \cite{decoh,Petruccione, Ullersma} reads:
\ben\label{H}
\hat H=\frac{\hat P^2}{2M}+\frac{M\Omega^2\hat X^2}{2}+\sum_{k=1}^N\left(\frac{\hat p_k^2}{2m_k}
+\frac{m_k\omega_k^2\hat x_k^2}{2}\right)+\hat X\sum_{k=1}^N C_k\hat x_k,
\een
where $\hat X, \hat P$ are the position and momentum of the central oscillator of mass M and frequency $\Omega$, $\hat x_k, \hat p_k$ are the positions and momenta 
of the bath oscillators, each with mass $m_k$ and frequency $\omega_k$, and $C_k$ are the coupling constants. This model can be in principle solved explicitly either directly \cite{Ullersma}
or using Wigner functions \cite{HaakeReibold}. However, as unlike in the standard treatments we are interested here not merely in the reduced state of the central oscillator alone,
but in the joint state of the central and a part of bath oscillators, the mentioned exact methods do not produce manageable solutions. As already stated in the Introduction,
the standard master equation methods are of no use either, since we are primarily interested in the influence of the system on the environment and not the other way around.
Following this "inverted" logic,  one can try to eliminate, at least in the first approximation, the recoil on the system due to the environment (apart from
the renormalization of the frequency).  This suggest a greatly simplifying assumption of a massive central system \cite{ZurekQBM,RoncagliaPaz}, which we will adopt.
One can then use a non-adiabatic version of the Born-Oppenheimer (Non-Born-Oppenheimer, NBO) approximation (see e.g. \cite{NBO}), where the central system evolves unperturbed, according to 
its self-Hamiltonian $\hat H_S=\hat P^2/(2M)+M\Omega^2\hat X^2/2$ (with the renormalized frequency $\Omega^2\equiv\Omega^2_{bare}-\sum_k C_k^2/(2m_k\omega_k^2)$) 
and the environment follows this evolution in the following way. The system propagator $K_t(X;X_0)\equiv\langle X|e^{-i\hat H_S t/\hbar}|X_0\rangle$ is rewritten
with the help of the classical trajectory $X(t;X_0)$, starting at $t=0$ at $X_0$ and reaching $X$ at time $t$ (as it is well known for the oscillator this semi-classical 
approximation is exact; see e.g. \cite{path}) and this trajectory acts as a classical driving force for the environment through the coupling term, leading to a controlled evolution:
\ben\label{drive}
i\hbar \frac{\partial}{\partial t}\ket{\psi_{E}(t)}=\hat H_{E}\left(X(t;X_0)\right)\ket{\psi_{E}(t)},
\een
where $\hat H_{E}\left(X(t;X_0)\right)\equiv \sum_{k=1}^N\left[\hat p_k^2/(2m_k)
+m_k\omega_k^2\hat x_k^2/2\right]+X(t;X_0)\sum_{k=1}^N C_k\hat x_k$. The full system-environment state is then constructed using the Born-Oppenheimer type of an  
ansatz:
\ben\label{final}
\Psi_{S:E}^{NBO}(X,\mathbf x)=\int dX_0 \phi_{S0}(X_0) \langle X|e^{-i\hat H_{S}t/\hbar}|X_0\rangle \langle\mathbf x|\hat U_{E}(X(t;X_0))\ket{\psi_{E0}},
\een 
with $\ket {\phi_{S0}}, \ket{\psi_{E0}}$ being the initial states of the system and the environment respectively, and $\hat U_{E}(X(t;X_0))$ is a solution of (\ref{drive}).

From the type of the coupling in (\ref{H}) and the analysis of the previous Section it follows that the candidates for the pointer states will be related to 
the position eigenstates. Hence, initial states of the system with large coherences in the position are of the greatest interest and for the purpose of this study
we choose an initially momentum-squeezed  ground state as the initial state of the system (studies of the
initially position-squeezed state from \cite{EPL} suggest that there is no SBS formation; cf. \cite{RoncagliaPaz}). 
This reduces the classical trajectories to $X(t;X_0)=X_0 \cos \Omega t$ \cite{ZurekQBM,RoncagliaPaz}
and in this case the evolution (\ref{final}) can be formally re-written using \cite{EPL}:
\ben\label{USE2}
\hat U_{S:E}(t)=\int dX_0 e^{-i\hat H_St/\hbar}\ket{X_0}\bra{X_0}\otimes \hat U_{E}(X_0\cos\Omega t).
\een
The driven evolution of the environment is easily solved and gives \cite{EPL}:
\ben\label{USE}
\hat U_{E}(X_0\cos\Omega t)=\bigotimes_{k=1}^N e^{i\zeta_k(t)X_0^2}e^{-i\sum_k\hat H_k t/\hbar}\hat D\left(\alpha_k(t)X_0\right)\equiv\bigotimes_{k=1}^N \hat U_k(X_0;t),
\een
where $\zeta_k(t)$ is a phase factor (unimportant for our considerations), $\hat H_k\equiv\hat p_k^2/(2m_k)+m_k\omega_k^2\hat x_k^2/2$, 
$\hat D(\alpha)\equiv e^{\alpha \hat a^\dagger-\alpha^*\hat a}$ is the displacement operator, and:
\ben
\alpha_k(t)\equiv -\frac{C_k}{2\sqrt{2 \hbar m_k\omega_k}}\left[\frac{e^{i(\omega_k+\Omega)t}-1}{\omega_k+\Omega}+
\frac{e^{i(\omega_k-\Omega)t}-1}{\omega_k-\Omega}\right]\label{ak}.
\een

The evolution (\ref{USE}) is formally a controlled-unitary type (\ref{CU}), in which the environment evolves accordingly to the initial position $X_0$ of the central oscillator. 
To study our central object---the partially reduced state (\ref{SfE}), one should be careful with the integral in (\ref{USE2}) as not to loose the diagonal part (\ref{diag}). We rewrite 
the integral using a finite division of the real line of $X_0$ into intervals $\{\Delta_i\}$ with $\ket{X_0} \bra{X_0}$ replaced by orthogonal projectors 
$\hat \Pi_{\Delta}$ on the intervals $\Delta$ (continuous distribution of $X_0$ is recovered in the limit of these divisions see e.g. \cite{GalindoPascual}). The partially traced state then reads:
\ben
&&\label{mama} \varrho_{S:fE}(t)=\sum_\Delta e^{-i\hat H_{S}t/\hbar}\hat \Pi_\Delta \ket{\phi_0}\bra{\phi_0} \hat \Pi_\Delta e^{i\hat H_{S}t/\hbar}\bigotimes_{k=1}^{fN} \varrho_k(X_\Delta;t)\\
&& +\sum_{\Delta\ne\Delta'} \Gamma_{X_\Delta,X_{\Delta'}}(t) e^{-i\hat H_St/\hbar}\hat \Pi_\Delta \ket{\phi_0}\bra{\phi_0} \hat \Pi_{\Delta'}e^{i\hat H_St/\hbar}
\otimes\bigotimes_{k=1}^{fN} \hat U_k(X_\Delta;t) \varrho_{0k}\hat U_k(X_{\Delta'};t)^\dagger,\nonumber
\een
where  $X_\Delta$ is some position within an interval $\Delta$,
\ben\label{rhoX}
\varrho_k(X_0;t)\equiv \hat U_k(X_0;t) \varrho_{0k}\hat U_k(X_0;t)^\dagger
\een
are the system-dependent states of the environments (\ref{Estates}), and: 
\ben
\Gamma_{X_0,X_0'}(t)\equiv \prod_{k\in (1-f)E}tr\left[\hat U_k(X_0;t) \varrho_{0k}\hat U_k(X_0';t)^\dagger\right]\equiv\prod_{k\in (1-f)E}\Gamma^{(k)}_{X_0,X_0'}(t)\label{G},
\een
is the decoherence factor. Following the general procedure of Section \ref{check}, one has to calculate it together with the generalized overlap  
(\ref{go}) for the states (\ref{rhoX}): $B^{(k)}_{X_0,X_0'}(t)\equiv B(\varrho_k(X_0;t), \varrho_k(X_0';t))$ as they serve as the indicator functions for the formation of
spectrum broadcast structures; cf. (\ref{indic}). Assuming the environment oscillators are initially in the thermal states with the same temperature,
the form of the decoherence factor has been known \cite{Petruccione}: 
\ben
&&\label{eq:dec} -\log | \Gamma^{(k)}_{X_0,X_0'}(t)  |  = \frac{(X_0-X_0')^2}{2}|\alpha_k(t)|^2\text{coth}\left(\tau_T\omega_k\right)=\\ 
&&\nonumber\frac{\left|X_0-X_0'\right|^2C_k^2\omega_k\coth \left(\tau_T\omega_k\right)}{4 \hbar m_k(\omega_k^2 - \Omega^2)^2}
\bigg[ \left(\cos \omega_k t - \cos \Omega t \right)^2 - \left(\sin \omega_k t - \frac{\Omega}{\omega_k}\sin \Omega t \right)^2 \bigg] \\ 
&& \equiv \frac{\left|X_0-X_0'\right|^2}{2}  f^{\Gamma}_{T}(t;\omega_k) \\ \nonumber  
\een
where $\tau_T\equiv\hbar/(2 k_BT)$ is the thermal time and we have introduced a function $f^{\Gamma}_{T}(t;\omega_k)\equiv |\alpha_k(t)|^2\text{coth}\left(\tau_T\omega_k\right)$ for a later convenience. 
The generalized overlap, in turn, for thermal (and also more general Gaussian) environments has been obtained in \cite{EPL}:
\ben\label{eq:ort}
&&\nonumber -\log  B^{(k)}_{X_0,X_0'}(t)  = \frac{(X_0-X_0')^2}{2}|\alpha_k(t)|^2\text{tanh}\left(\tau_T\omega_k\right) \\
&&\equiv \frac{\left|X_0-X_0'\right|^2}{2}  f^{B}_{T}(t;\omega_k). 
\een
We note that the factor $\text{coth}\left(\tau_T \omega_k\right)$ appearing in the decoherence factor is related to the mean initial energy 
of the environmental oscillators at temperature $T$, $\text{coth}\left(\tau_T \omega_k\right)=\langle E(\omega,T)\rangle/E_0(\omega)$, where $E_0(\omega)\equiv\hbar\omega/2$ is the zero-point energy,
while $\text{tanh}(\tau_T \omega_k)$, appearing in the generalized overlap, is nothing else but the purity $tr(\varrho_{0k}^2)$  of the initial thermal state $\varrho_{0k}$, which in turn is related
to the linear entropy $S_{lin}(\varrho_{0k})=1-tr(\varrho_{0k}^2)$. Thus, the effectiveness of the decoherence depends on the initial energy of the environment, while 
information accumulation on its purity.

To proceed further with the analysis, one has to specify the environment. The standard procedure \cite{Schlosshauer,decoh,Petruccione,Ullersma,ZurekQBM,RoncagliaPaz} is to pass to a continuum limit 
of frequencies $\omega_k$ and encode the properties of the environment in a specific continuous approximation to 
the spectral density function $J(\omega)=\sum_k C_k^2/(2m_k\omega_k)\delta(\omega-\omega_k)$ (e.g. in \cite{ZurekQBM,RoncagliaPaz} the Ohmic spectral density has been chosen).
In contrast, in  \cite{EPL} a somewhat different idea has been put forward: To keep the environment discrete but random, with frequencies $\omega_k$
chosen from some given ensemble. Randomness is needed to effectively induce decoherence in the spirit of \cite{ZurekSpins}, as the environment 
remains finite-dimensional. For definiteness' sake the simplest case has been studied, where the frequencies $\omega_k$ are independently, identically 
distributed (i.i.d.) with a uniform distribution over a finite interval $[\omega_L, \omega_U]$. The interval is chosen so that the environment is off-resonant 
(cf. (\ref{eq:dec},\ref{eq:ort})), to avoid decohereing of the system by a single environment, and "fast":
\ben\label{fast}
\omega_U,\omega_L\gg\Omega.
\een
This choice of the environment may be considered as a direct, "mechanstic", as opposed to the usual field, treatment of the environment:
The bath is a collection of identical mechanical oscillators with masses $m_k$ and random frequencies $\omega_k$. 
It leads to a complication in the study of the conditions (\ref{indic}) as from (\ref{G}, \ref{eq:dec}) and (\ref{Bmac}, \ref{eq:ort}) the macroscopic indicator functions,
associated with the traced over part of the environment $(1-f)E$ and an observed macfraction $mac$ respectively:
\ben
&&\left|\Gamma_{X_0,X_0'}(t)\right|=\exp\left[-\frac{\left|X_0-X_0'\right|^2}{2}\sum_{k\in(1-f)E} f^{\Gamma}_{T}(t;\omega_k)\right],\label{Gmac}\\ 
&&B_{X_0,X_0'}(t)=\exp\left[-\frac{\left|X_0-X_0'\right|^2}{2}\sum_{k\in mac} f^{B}_{T}(t;\omega_k)\right]\label{Bmac2}
\een
become almost periodic functions of time. Previously \cite{EPL} those functions were studied only numerically, indicating that indeed there is a parameter regime
that the time averages over very long times of the above (non-negative) functions simultaneously vanish, indicating small typical fluctuations above zero.
This in turn, implies that the partially traced state (\ref{mama}) approaches dynamical spectrum broadcast form with respect to the initial position $X_0$:
\ben\label{sbsqbm}
\varrho_{S:fE}(t)&\approx&\int dX_0 \left|\langle X_0|\phi_0\rangle\right|^2 e^{-i\hat H_St}\ket{X_0}\bra{X_0} e^{i\hat H_St}\otimes\nonumber\\
&\otimes&\varrho_{mac_1}(X_0;t)\otimes \cdots\otimes \varrho_{mac_{\mathcal M}}(X_0;t),
\een
with $\varrho_{mac_k}(X_0;t)$ having non-overlapping supports. The dynamical character of the above structure 
is manifest in the time dependence of all the appearing states: The pointer basis is rotating according to the system self-Hamiltonian
(thanks to the recoil-free assumption) $\ket{X(t)}\equiv e^{-i\hat H_St}\ket{X_0}$ and this motion modulates the evolution of the
environment in such a way that an instantenous SBS is formed at any moment. In a sense, the environment encodes the motion of the
central oscillator. 
The main parameters this process depends on through (\ref{Gmac},\ref{Bmac2}) are time, temperature, 
separation $\Delta X_0\equiv|X_0-X_0'|$ \cite{Photonics}, and macrofraction size $N_{mac}$. Trade-offs between them dictate if and when the structure (\ref{sbsqbm}) will be formed.
In what follows we study this behavior analytically, assuming large macrofraction size $N_{max}$.

\section{Analytical estimates of the SBS formation}
As mentioned in the previous Section, we are working with  random  environments with i.i.d. frequencies $\omega_k$ 
with some distribution  $P(\omega)$.
As a consequence, the functions $f^{B}_{T}(t;\omega_k) $ and $f^{\Gamma}_{T}(t;\omega_k)$, appearing in the SBS indicator functions 
(\ref{Gmac},\ref{Bmac2}),  also become i.i.d. random variables for a fixed time $t$ and temperature $T$. 
Analytical study of their sums over a macrofraction $\sum_{k=1}^{N_{mac}} f^{\Gamma,B}_{T}(t;\omega_k)$ (we assume for simplicity that
both the unobserved macrofraction $(1-f)E$ as well as each of the observed ones have the same size $N_{mac}$) is possible in the limit 
of a large macrofraction size $N_{mac} \rightarrow \infty$ using the Law of Large Numbers (LLN) \cite{LLN}. This will be our main tool. It states (in its strong form)
that the macrofraction averages $1/N_{mac}\sum_{k=1}^{N_{mac}} f^{\Gamma,B}_{T}(t;\omega_k)$ converge almost surely, i.e. with probability one,
to their expectation values:
\ben
\frac{1}{N_{mac}}\sum_{k=1}^{N_{mac}} f_{T}(t;\omega_k)\stackrel{a.s.}{\rightarrow}
\int d\omega P(\omega) f_{T}(t;\omega)\equiv\left\langle \left\langle f_{T}(t;\omega)  \right\rangle\right\rangle 
\een
(we will be neglecting the superscripts ${\Gamma,B}$ unless it leads to a confusion)
and, according to the large deviation theory, the probability of error is exponentially small in $N_{mac}$ with the rate governed by the, so called, 
rate function (which we will not be interested in here, only assuming that it exists and is non-zero).
This allows us to approximate the sums $\sum_{k=1}^{N_{mac}} f_{T}(t;\omega_k)$ with $N_{mac} \left\langle \left\langle f_{T}(t;\omega)  \right\rangle\right\rangle$.

We note that the invocation of  LLN is, in this context, effectively equivalent to the continuous limit for the macrofractions of the environment with $P(\omega)$
determining the spectral density. In other words, we divide the environment into fractions of such a large size that the LLN may be applied. 
Following our approach, explained in the previous Section, instead of the standard spectral densities, such as e.g. Ohmic, we will use here
a much simpler, uniform probability distribution over an interval $[\omega_L,\omega_U]$ due to an ease of analysis:
\ben
\label{eq:lawoflarge}
\left\langle \left\langle f_{T}(t;\omega)  \right\rangle\right\rangle  =   \frac{1}{\Delta \omega} \int^{\omega_U}_{\omega_L} d \omega f_{T}(t;\omega) , 
\een  
where  $\Delta \omega = \omega_U - \omega_L$. In what follows we analyze the short- and long-time behavior of this expression in the limits of
high and low temperature. This will enable us to estimate the macrofraction size $N_{mac}$ needed in order for the functions (\ref{Gmac},\ref{Bmac2}) 
to attain asymptotically values close to zero within a given error as well as give the timescales of their initial decays, observed numerically in \cite{EPL}.

\subsection{Low temperature}
Let us first assume that the temperature is so low, that the associated thermal energy is much lower than the lowest oscillator energy: $ k_B T\ll \hbar \omega_L$.
Then in the leading order the temperature dependence can be neglected $\coth(\hbar \omega_k/2 k_B T) \approx \tanh (\hbar \omega_k/2 k_B T) \approx 1$  
and the behavior of decoherence and orthogonalization becomes identical:
\ben
\label{eq:2}
f^{\Gamma}_{T}(t;\omega_k)  \approx f^{B}_{T}(t;\omega_k)  \approx |\alpha_k(t)|^2\equiv f_0(t;\omega_k)
\een
with $\alpha_k(t)$ given by (\ref{ak}).
The calculation of the ensemble mean of $f_0(t; \omega_k)$ is rather lengthy and is presented in \ref{ap:low}, with 
an assumption that the interaction strengths $C_k$ obey (cf. \cite{ZurekQBM,RoncagliaPaz}) $C_k = 2 \sqrt{(M m_k \bar{\gamma}_0)/\pi}$,
with $\bar{\gamma}_0$ a constant.

First, we are interested in the short-time behavior, valid for times much shorter than the shortest timescale of the full Hamiltonian, which in this case is 
$t\ll\omega_U^{-1}$ (we recall that we assume $\Omega$ to be much lower than the environmental frequencies in order to be in the off-resonant regime, so that collections
rather than individual environments matter).
By expanding the expression for $\left\langle \left\langle f_0(t;\omega)   \right\rangle\right\rangle$ in power series with respect to time, 
we find after a tedious calculation that (for details see \ref{ap:low} eq. (\ref{ap:eq:ltst})):   
\ben
\left\langle \left\langle f_0(t;\omega)   \right\rangle\right\rangle  = \frac{2 M \bar{\gamma}_0}{ \hbar \pi \Delta \omega}  \log \left( \frac{\omega_U}{\omega_L} \right)t^2+O(t^4),
\een
which immediately implies that the initial behavior of both the decoherence and the orthogonalization factors is a Gaussian decay (c.f. (\ref{eq:dec}, \ref{eq:ort})):
\ben
\label{eq:stbl}
&&|\Gamma_{X_0,X_0'}(t)| \approx  B_{X_0,X_0'}(t) \approx \exp \left[ -N_{mac} \left(\frac{t}{\tau_0}\right)^2 \right] ,
\een
with a common timescale: 
\ben
\frac{\tau_0}{\sqrt{N_{mac}}}, \ \tau_0= \frac{\hbar \pi \Delta \omega}{\Delta X_0 M \bar{\gamma}_0}\log^{-1} \left(  1+\frac{\Delta\omega}{\omega_L}\right). 
\een
We note that it depends on the macrofraction size and the separation through the product $\Delta X_0\sqrt{N_{mac}}$. Thus, in order to keep the same time-scale for small separations 
the macrofraction size should increase quadratically with decreasing separation.  

The initial Gaussian decay (\ref{eq:stbl}) by no means guarantees that the functions will stay close to zero with negligible fluctuations---revivals are possible, as has been shown in \cite{EPL}.
Thus a long-time analysis is needed, governed in our case by the condition $t\gg 1/(\omega_L - \Omega)\approx 1/\omega_L$ as $\Omega\ll\omega_L$ 
The detailed calculation is tedious and is given in the \ref{ap:low}, eq. (\ref{ap:eq:ltlt}). The result reads:
\ben\label{lowTas}
\label{eq:ltltdn}
\left\langle \left\langle  f_0(t;\omega)    \right\rangle\right\rangle = \frac{2 M \bar{\gamma}_0}{\hbar \pi \Delta \omega}\left(A_0 \cos^2 (\Omega t) + B_0 \right),
\een
where:
\ben
&&A_0\equiv-\frac{1}{2\Omega^2}\bigg( 2 \log \frac{\omega_U}{\omega_L} - \log \frac{\omega_U^2-\Omega^2}{\omega_L^2-\Omega^2} \bigg),\\
&&B_0\equiv \frac{1}{\omega_L^2-\Omega^2}-\frac{1}{\omega_U^2-\Omega^2} - A_0.\\
\een
Interestingly, for large times the mean has an oscillatory part with the system frequency $\Omega$, but for fast environments (\ref{fast}) 
this part is vanishingly small as $A_0\approx 0$. The above formulas allow us to solve
a very important problem in the context of SBS: How big should be choose macrofractions in order to get decoherence and 
orthogonalization with a prescribed error $\epsilon$ (common in the low $T$ limit for both functions): 
\ben\label{error}
|\Gamma_{X_0,X_0'}(t)|, B_{X_0,X_0'}(t) < \epsilon.
\een
This in turn will determine the trace norm distance of the actual state $\varrho_{S:fE}(t)$ to the spectrum broadcast form.    
From (\ref{lowTas}) and (\ref{Gmac},\ref{Bmac2}) we immediately obtain that if: 
\ben\label{result}
\Delta X_0^2 N_{mac} > \frac{\hbar \pi\Delta\omega}{ M\bar\gamma_0 B_0}\log  \frac{1}{\epsilon}\approx 
\frac{\hbar \pi\omega_U^2\omega_L^2}{ M\bar\gamma_0 \left(\omega_U+\omega_L\right)}\log  \frac{1}{\epsilon},
\een      
then the functions will be bounded by (\ref{error}) for all times $t\gg 1/(\omega_L-\Omega)$. This result can be 
treated as an analytical proof of SBS formation in the studied regime. Similarly to the short-time decay (\ref{eq:stbl}), the
asymptotic behavior of $|\Gamma_{X_0,X_0'}(t)|, B_{X_0,X_0'}(t)$ is governed by the product $\Delta X_0^2 N_{mac}$,
so that the increase of the macrofraction size is quadratic with decreasing the spatial resolution of the SBS. 
This finite spatial resolution of the SBS for a given error level and a macrofraction size is a manifestation of the 
"macroscopic objectivity" idea, introduced in \cite{Photonics} for simplified models of QBM.
Namely, for a given tolerance $\epsilon$ and a macrofraction size, the objective state of the system
appear only on the length scales greater than ones given by (\ref{result}).

\subsection{High temperature}
Here we consider the opposite situation of a hot environment: 
$k_B T \gg \hbar \omega_U $. Intuitively, a formation of the SBS should be quite compromized now, as high temperature, while increasing the 
decoherence power of the environment through the increase of its energy appearing in (\ref{eq:dec}), decreases its information capacity, by decreasing
the purity, on which depends the orthogonalization factor (\ref{eq:ort}).
Indeed, this is what we show below. In the leading order $\tanh (\tau_T\omega) =[\coth (\tau_T\omega)]^{-1} \approx \tau_T\omega $ and 
(\ref{eq:dec}) and (\ref{eq:ort}) read:
\ben
&&f^{\Gamma}_{T}(t;\omega_k) \approx \frac{1}{\tau_T \omega_k} |\alpha_k(t)|^2, \\ 
&&f^{B}_{T}(t;\omega_k) \approx \tau_T \omega_k |\alpha_k(t)|^2.
\een  
The relevant means (\ref{eq:lawoflarge}) can be calculated analytically again; see \ref{ap:highd} and \ref{ap:higho}. 
For short time-scales $t\ll\omega_U^{-1}$ we obtain the following behavior (for the details see \ref{ap:high}, eq. (\ref{ap:eq:htdst}), (\ref{ap:eq:htost})):
\ben
\label{eq:dsh}
\left\langle \left\langle f^{\Gamma}_{T}(t;\omega) \right\rangle\right\rangle  =  \frac{2 M \bar{\gamma}_0}{\hbar \pi  \omega_L \omega_U \tau_T}t^2 + O(t^4), \\
\label{eq:osh}
\left\langle \left\langle f^{B}_{T}(t;\omega) \right\rangle\right\rangle  =   \frac{2 M \bar{\gamma}_0 \tau_T}{\hbar \pi}\tau_T t^2 + O(t^4),
\een
resulting again in the initial Gaussian decay:
\ben
&&|\Gamma_{X,X'}(t)|  \approx \exp \left[ -N_{mac} \left(\frac{t}{\tau_{dec}} \right)^2 \right] \\ 
&&B_{X,X'}(t) \approx \exp \left[ -N_{mac} \left(\frac{t}{\tau_{ort}} \right)^2 \right].
\een
However, this time the timescales are different. For the decoherence one obtains (cf. (\ref{ap:eq:htdst})) 
\ben
\label{eq:htdst}
&&\frac{\tau_{dec}}{\sqrt{N_{mac}}}, \; \; \tau_{dec} = \tau_T\frac{\hbar \pi  \omega_L \omega_U}{\Delta X_0 M \bar{\gamma}_0},
\een
whereas for generalized overlap (cf. (\ref{ap:eq:htost})) the characteristic time is:
\ben
\label{eq:htost}
&&\frac{\tau_{ort}}{\sqrt{N_{mac}}}, \; \; \tau_{ort} = \tau_T^{-1}\frac{\hbar \pi}{\Delta X_0 M \bar{\gamma}_0 }.
\een
As one would expect, the key difference is in the temperature dependence through the the thermal time $\tau_T=\hbar/(2k_B T)$. While $\tau_{dec}$
decreases as $T^{-1}$ indicating faster decoherence with higher temperature, $\tau_{ort}\sim T$ so that it may even happen that the orthogonalization timescale $\tau_{dec}/\sqrt{N_{mac}}$
is larger than the validity of the short-time approximation $t\ll\omega_U^{-1}$. Keeping $\tau_{dec}/\sqrt{N_{mac}}<\omega_U^{-1}$ so that the short-time approximation, and hence the Gaussian decay,
is valid, puts a constraint on 
the temperature, the macrofraction size and the separation to be discriminated:
\ben
\frac{T}{\Delta X_0\sqrt{N_{mac}}}<\frac{M\bar\gamma_0}{2\pi k_B\omega_U}.
\een

To get some insight into possible revivals of the decoherence and orthogonalization factors, we perform long-time analysis. 
In \ref{ap:high} it is shown that for $t \gg 1/(\omega_L-\Omega)\approx1/\omega_L$ the asymptotic expression for 
$\left\langle \left\langle f^{\Gamma}_{T}(t;\omega)   \right\rangle\right\rangle$ reads:
\ben
\label{eq:htasd}
\left\langle \left\langle f^{\Gamma}_{T}(t;\omega)   \right\rangle\right\rangle = &&\frac{2 M \bar{\gamma}_0}{\hbar \pi  \Delta \omega\tau_T}\left( A_{\Gamma} \cos^2 (\Omega t) + B_{\Gamma}   \right) + O(t^{-1})
\een 
with:
\ben
&&A_{\Gamma} \equiv -\frac{1}{4\Omega^2}\bigg[\frac{\Delta\omega}{\omega_U \omega_L } + \frac{1}{2\Omega} \log \frac{(\omega_U+\Omega)(\omega_L-\Omega)}{(\omega_U-\Omega)(\omega_L+\Omega)}  \bigg], \nonumber \\
&&B_{\Gamma} \equiv \frac{1}{4\Omega^2}\bigg( \frac{\omega_L}{\omega_L^2-\Omega^2} - \frac{\omega_U}{\omega_U^2-\Omega^2} \bigg) - A_{\Gamma},\nonumber,
\een
while for generalized overlap it is:
\ben
\label{eq:htaso}
\left\langle \left\langle f^{B}_{T}(t;\omega)   \right\rangle\right\rangle =  &&\frac{2 M \bar{\gamma}_0 \tau_T}{ \hbar \pi \Delta \omega}\left(A_{B} \cos^2 (\Omega t) + B_{B}  \right) + O(t^{-1}),
\een
where:
\ben
&&A_{B} \equiv \frac{1}{2\Omega}\log\frac{(\omega_U-\Omega)(\omega_L+\Omega)}{(\omega_L-\Omega)(\omega_U+\Omega)} \\
&&B_{B} \equiv \frac{\omega_L}{\omega_L^2-\Omega^2} - \frac{\omega_U}{\omega_U^2-\Omega^2}. 
\een
We observe that unlike in the low $T$ regime, the decoherence asymptotic keeps oscillating with the system frequency $\Omega$ even
for fast environments (\ref{fast}) as $A_\Gamma\approx \Delta\omega/(4\Omega^2\omega_U\omega_L)$, while $A_B\approx 0$. We are now ready to solve the problem of the SBS formation in the high temperature regime: 
In a given temperature $T$, how big should be the macrofraction sizes to achieve decoherence and distinguishability, and hence the SBS, on a length-scale $\Delta X_0$
within given errors: 
\ben
\label{eq:hterror}
|\Gamma_{X_0,X_0'}(t)| < \epsilon_{dec}, \ B_{X_0,X_0'}(t) < \epsilon_{ort}?
\een
Eqs.  (\ref{eq:htasd}) and (\ref{eq:htaso}) give us the answer:
\ben
\label{eq:htltdn}
&&T \Delta X^2_0 N^{\Gamma}_{mac} > \frac{\hbar^2 \pi \Delta \omega}{2 M k_B \bar{\gamma}_0 B_\Gamma}   \log \frac{1}{\epsilon_{dec}} \approx \frac{\hbar^2 \pi \Omega^2 \omega_U\omega_L}{M k_B \bar{\gamma}_0 }  
 \log \frac{1}{\epsilon_{dec}},  \\
\label{eq:htlton}
&&\frac{\Delta X^2_0 N^{B}_{mac}}{T} > \frac{ 2 \pi k_B  \Delta \omega}{M \bar{\gamma}_0 B_B} \log \frac{1}{\epsilon_{ort}} \approx \frac{ 2 \pi k_B  \omega_U\omega_L}{M \bar{\gamma}_0} \log \frac{1}{\epsilon_{ort}},
\een
where $N^{\Gamma}_{mac}$ is the size of the traced-over part of the environment $(1-f)E$ and $N^{B}_{mac}$ is the size of (each of) the observed macrofraction. As predicted, keeping all other parameters fixed,
the observed macrofraction size in high temperature must be much larger than the unobserved one in order to come close to SBS. Indeed, from the above results those sizes scale like
the thermal-to-central-system energies:  
\ben
&&\frac{N^B_{mac}}{N^B_{mac}} > 2\left(\frac{k_B T}{\hbar \Omega} \right)^2 \frac{\log \epsilon_{ort}}{\log \epsilon_{dec}}
\een 
and the later factor is huge for the considered fast environments, since $k_BT\gg\hbar\omega_U\gg\hbar\Omega$.

\section{Conclusions}
We have studied the process of formation of the spectrum broadcast structures in Quantum Brownian Motion model,
continuing the research initiated in \cite{EPL}. Being interested in the information gained by the environment about the system, 
we have considered a rather non-standard limit of a massive central system (initially in the momentum squeezed state),
 and somewhat simplified  random environments with i.i.d. uniformly distributed frequencies. 
The use of the Law of Large Numbers, assuming the environment to be sufficiently large, 
allowed us to obtain analytical results on the spectrum broadcast structure formation.

In particular, we have investigated short-time behavior of decoherence and generalized overlap factors in low and high temperatures. 
In the low temperature regime, we have shown that both factors admit Gaussian decay with the same timescale, which depends on frequencies 
of unobserved environmental oscillators only. In the high temperature regime they also decay in a Gaussian way. However the resulting
timescales are different functions of temperature: the decoherence rate is proportional to the temperature, whereas the rate of decay of generalized overlap is inversely proportional to temperature. 
This explains in a quantitative way previous numerical simulations showing rapid decoherence and vanishing orthogonalization of remaining environmental states in the studied 
model with growing temperature for fixed number of environmental systems.

Long-time analysis gave us the efficiency of the spectrum broadcast structure formation in the sense of the required observed/unobserved macrofraction sizes
to obtain decoherence and the environmental state distinguishability within given errors. 
In low temperatures these sizes are equal, but, as one would expect, in high temperature they have the opposite temperature dependence
as hot environments decohere the central system efficiently but encode a vanishingly small amount of information due to high noise.

An obvious generalization of the present work would be an analysis of more standard environment models, e.g. the Ohmic one with a cut-off.
The resulting functions will be more complicated but we believe still analyzable, at least in certain approximate regimes, similar to the studied above.

\section{Acknowledgements}
We would like to thank  J. Wehr, P. Horodecki, and R. Horodecki for discussions and remarks. JKK acknowledges the financial support of 
the John Templeton Foundation through the grant ID \#56033. JT acknowledges
support of the Polish National Science Center by means of project no. 2015/16/T/ST2/00354 for the PhD thesis.

\newpage
\appendix
\section{Appendix Content}
\ref{ap:sci} is devoted to Sine and Cosine Integrals. In \ref{ap:scinot} we introduce notation for particular combinations of Sine and Cosine integrals that appear in formulas. In \ref{ap:scistb} and  \ref{ap:sciltb} formulas for short-time and long-time behavior of this functions are presented. \ref{ap:low} contains details of computations for low-temperature Quantum Brownian Motion. The high-temperature case is treated in \ref{ap:high} with \ref{ap:highd}, \ref{ap:higho} devoted to decoherence factor and generalized overlap respectively. 
\section{Sine and Cosine integrals}
\label{ap:sci}
\subsection{Notation}
\label{ap:scinot}
It will prove beneficial for the sake of clarity to introduce the following notation for combinations of Sine and Cosine integrals: 
\ben
&&F_{\text{Si}}(\pm,\pm,\pm,\pm) = [\pm 1,\pm 1,\pm 1,\pm 1] \cdot \nonumber \\  &&\left[\text{Si}\left((\omega_L - \Omega)t \right), \text{Si} \left( (\omega_U - \Omega)t \right),  \text{Si}\left((\omega_L + \Omega)t \right), \text{Si} \left( (\omega_U + \Omega)t \right) \right]^T \\
&&F_{\text{Ci}}(\pm,\pm,\pm,\pm) =  [\pm 1,\pm 1,\pm 1,\pm 1] \cdot\nonumber \\ &&\left[\text{Ci}\left((\omega_L - \Omega)t \right),  \text{Ci} \left( (\omega_U - \Omega)t \right), \text{Ci}\left((\omega_L + \Omega)t  \right), \text{Ci} \left( (\omega_U + \Omega)t \right) \right]^T,
\een
where $[\pm 1, \ldots \pm 1]$ is a vector, $\cdot$ denotes vector product and $T$ stands for transposition. The argument of $F_{\text{Si}}(\pm,\pm,\pm,\pm) , F_{\text{Ci}}(\pm,\pm,\pm,\pm)$ specifies signs' pattern of functions e.g.:
\ben
F_{\text{Si}}(+,-,+,-) = \text{Si}\left((\omega_L - \Omega)t \right) - \text{Si} \left( (\omega_U - \Omega)t \right) +  \text{Si}\left((\omega_L + \Omega)t \right) - \text{Si} \left( (\omega_U + \Omega)t \right) \nonumber
\een
\subsection{Short-time behavior}
\label{ap:scistb}
In the short time regime, i.e. for $t \ll \omega_U^{-1}$, we can approximate relevant functions as follows:
\ben
&&F_{\text{Si}}(+,-,+,-) = 2(\omega_L-\omega_U)t + \frac{t^3}{9}(\omega_U^3-\omega_L^3+3\Omega^2\omega_U-3\Omega^2\omega_L) +O(t^5)  \nonumber \\
&&F_{\text{Si}}(+,-,-,+) =\frac{t^3}{3}\Omega(\omega_L^2-\omega_U^2) +O(t^5) \nonumber  \\
&&F_{\text{Ci}}(+,-,+,-) =  \log \frac{\omega_L^2-\omega^2}{\omega_U^2-\omega^2} + \frac{1}{2} (\omega_U^2-\omega_L^2)t^2 +o(t^4) \nonumber  \\
&&F_{\text{Ci}}(+,-,-,+) = \log\frac{(\omega_L-\Omega)(\omega_U+\Omega)}{(\omega_L+\Omega)(\omega_U-\Omega)} + \Omega(\omega_L-\omega_U)t^2 +O(t^4),
\een
\subsection{Long-time behavior}
On the other hand, the asymptotic of relevant functions is given by: 
\label{ap:sciltb}
\ben
tF_{\text{Si}}(+,-,+,-) =  &&2 \bigg( \omega_U \cu \com + \Omega \su \som - \nonumber \\ &&\omega_L \cl \com - \Omega \sl \som \bigg) + O(t^{-1}) \nonumber \\
tF_{\text{Si}}(+,-,-,+) =  &&2 \bigg( \omega_U \su \som + \Omega \cu \com - \nonumber \\ &&\omega_L \sl \som - \Omega \cl \com \bigg) + O(t^{-1}) \nonumber \\
tF_{\text{Ci}}(+,-,+,-) = &&2 \bigg( \omega_L \sl \com - \Omega \sl \som - \nonumber \\ &&\omega_U \su \com + \Omega \cu \som \bigg) + O(t^{-1}) \nonumber \\
tF_{\text{Ci}}(+,-,-,+) = &&2 \bigg( \Omega \sl \com - \omega_L \cl \som   + \nonumber \\ &&\omega_U \cu \som - \Omega \su \com \bigg) + O(t^{-1}). \nonumber \\
\een

\section{Low temperature}
\label{ap:low}
Here we present details of computing $\left\langle \left\langle f^{\Gamma}_{T}(t;\omega)  \right\rangle\right\rangle$ and $\left\langle \left\langle f^{B}_{T}(t;\omega)  \right\rangle\right\rangle$ in the low temperature regime. As was already mentioned in the main text, in this case $ \left\langle \left\langle f^{\Gamma}_{T}(t;\omega)  \right\rangle\right\rangle \approx \left\langle \left\langle f^{B}_{T}(t;\omega)  \right\rangle\right\rangle \approx \left\langle \left\langle f_0(t;\omega)  \right\rangle\right\rangle$, 
\ben
\label{ap:eq:fint}
\left\langle \left\langle f_0(t;\omega)  \right\rangle\right\rangle = && \frac{2 M \bar{\gamma}_0}{ \hbar \pi \Delta \omega} \int^{\omega_U}_{\omega_L} \frac{\omega_k}{(\omega_k^2 - \Omega^2)^2}\bigg(\left( 1 -  \cos^2 \Omega t \right) + \frac{\Omega^2}{\omega^2} \left( 1 + \cos^2 \Omega t  \right) \nonumber \\ &&- 2 \cos \Omega t \cos \omega t - \frac{2 \Omega}{\omega_k} \sin \Omega t \sin \omega t \bigg)  = \nonumber \\ && \frac{2 M \bar{\gamma}_0}{ \hbar \pi \Delta \omega}\left( I_1 + I_2 -2 I_3 -2 I_4 \right) ,
\een
where the mean consists of four integrals. Results of each integration are given by:
\ben
&&I_1=\int^{\omega_U}_{\omega_L} \frac{\omega}{(\omega^2-\Omega^2)} (1+ \cos^2 (\Omega t)) = \\ \nonumber &&\frac{1}{2}\bigg(\frac{1}{\omega_L^2-\Omega^2}-\frac{1}{\omega_U^2-\Omega^2} \bigg)(1+ \cos^2 (\Omega t)) \\ \nonumber \\ 
&&I_2=\int^{\omega_U}_{\omega_L} \frac{\Omega^2}{\omega(\omega^2-\Omega^2)}(1- \cos^2 (\Omega t))= \\ \nonumber
&&\left[\frac{1}{4\Omega^2}\bigg( 4 \log \frac{\omega_U}{\omega_L} - 2 \log \frac{\omega_U^2-\Omega^2}{\omega_L^2-\Omega^2} \bigg) +\frac{1}{2(\omega_L^2-\Omega^2)}-\frac{1}{2(\omega_U^2-\Omega^2)}\right](1- \cos^2 (\Omega t)) \\ \nonumber \\
&&I_3=\int^{\omega_U}_{\omega_L} d \omega \frac{\omega}{(\omega^2 - \Omega^2)^2} \cos \omega t \cos \Omega t =  \\ \nonumber &&\frac{1}{4 \Omega} \cos \Omega t \left[ \frac{2 \Omega \cos \omega_Lt}{\omega_L^2 - \Omega^2} - \frac{2 \Omega \cos \omega_Ut}{\omega_U^2 - \Omega^2} \right. + \\ \nonumber &&  t \cos \Omega t  F_{\text{Si}}(+,-,-,+) + t \sin \Omega t F_{\text{Ci}}(+,-,+,-)   \bigg] \\ \nonumber \\
&&I_4=\int^{\omega_U}_{\omega_L} d \omega \frac{\Omega}{(\omega^2 - \Omega^2)^2} \sin \omega t \sin \Omega t =  \\ \nonumber &&  \frac{1}{4 \Omega} \sin \Omega t \bigg \{ \frac{2 \omega_L\sin \omega_L t} {\omega_L^2 - \Omega^2} - \frac{2 \omega_U\sin \omega_U t} {\omega_U^2 - \Omega^2} + \\ &&  t \left[ \cos (\Omega t) F_{\text{Ci}}(-,+,-,+) \nonumber + \sin (\Omega t) F_{\text{Si}}(+,-,-,+) \right] - \nonumber \\ \nonumber
&&\Omega^{-1} \left[F_{\text{Si}}(-,+,+,-) - F_{\text{Ci}}(-,+,-,+)\right]  \bigg \}.
\een
To investigate the short-time behavior of $\left\langle \left\langle f_0(t;\omega)   \right\rangle\right\rangle$ we expand the above expressions up to the second order in time. This is a good approximation for $t \ll \omega_U^{-1}$. As a result we obtain 
\ben
&&I_{1} =  \\ \nonumber &&\left[\frac{1}{4\Omega^2}\bigg( 4 \log \frac{\omega_U}{\omega_L} - 2 \log \frac{\omega_U^2-\Omega^2}{\omega_L^2-\Omega^2} \bigg) +\frac{1}{2(\omega_L^2-\Omega^2)}-\frac{1}{2(\omega_U^2-\Omega^2)}\right]\Omega^2t^2 + O(t^4)  \\ \nonumber \\ 
&&I_2 = \bigg(\frac{1}{\omega_L^2-\Omega^2}-\frac{1}{\omega_U^2-\Omega^2} \bigg) \bigg(2 -\Omega^2t^2 \bigg) + O(t^4)   \\ \nonumber \\ 
&&I_3 = \frac{1}{4 \Omega}\bigg[\frac{2 \Omega}{\omega_L^2 - \Omega^2}\bigg(1-\frac{\omega_L^2  t^2}{2} -\frac{\Omega^2  t^2}{2}  \bigg)  
 -  \\ &&  \frac{2 \Omega}{\omega_U^2 - \Omega^2} \bigg(1-\frac{\omega_U^2  t^2}{2} -\frac{\Omega^2  t^2}{2} 
\bigg) \nonumber     
+ \Omega t^2 
  \log \frac{\omega_L^2-\Omega^2}{\omega_U^2-\Omega^2}   \bigg] + O(t^4) 
	\\ \nonumber \\
&& I_4 =  \frac{1}{4 \Omega}\bigg ( \frac{2 \omega_L \omega_L t} {\omega_L^2 - \Omega^2} \Omega \omega_L t^2 
\ \frac{2 \omega_U \omega_U t} {\omega_U^2 - \Omega^2}  \Omega \omega_U t^2  
\bigg) + O(t^4).
\een
As a result, the expression for the mean valid in the short-time regime is  
\ben
\label{ap:eq:ltst}
\left\langle \left\langle f_0(t;\omega)   \right\rangle\right\rangle =\frac{2 M \bar{\gamma}_0}{ \hbar \pi \Delta \omega} \log\left(\frac{\omega_U}{\omega_L}\right)t^2 + O(t^4),
\een
what leads to (\ref{eq:stbl}) of text main text.

On the other hand, using asymptotic formulas from \ref{ap:sciltb} one can show that for $t\gg (\omega_L - \Omega)^{-1}$ $I_3 \approx 0$ and $I_4 \approx 0$ so the only relevant terms are $I_1$ and $I_2$, what results in the following expression for long-time behavior of the mean
\ben
\label{ap:eq:ltlt}
\left\langle \left\langle f_0(t;\omega)    \right\rangle\right\rangle = \frac{2 M \bar{\gamma}_0}{\hbar \pi \Delta \omega}\left( A_0 \cos^2 (\Omega t) + B_0\right) + O(t^{-1}),
\een
where
\ben
&&A_0\equiv-\frac{1}{2\Omega^2}\bigg( 2 \log \frac{\omega_U}{\omega_L} - \log \frac{\omega_U^2-\Omega^2}{\omega_L^2-\Omega^2} \bigg), \nonumber \\
&&B_0\equiv \frac{1}{\omega_L^2-\Omega^2}-\frac{1}{\omega_U^2-\Omega^2} - A_0\nonumber
\een
Minimization of (\ref{ap:eq:ltlt}) is straightforward, yields minimal value $\frac{2 M \bar{\gamma}_0 B_0}{\hbar \pi \Delta \omega}$, what was used to derive formula (\ref{eq:ltltdn}) from text main text.

\section{High temperature}
\label{ap:high}
In the case of high temperature one can approximate functions appearing in decoherence factor and generalized overlap as   
\ben
&&f^{\Gamma}_{T}(t;\omega_k) \approx \frac{1}{\tau_T \omega_k} f_0(t;\omega_k) ,  \;\;\; f^{B}_{T}(t;\omega_k) \approx \tau_T \omega_k f_0(t;\omega_k) . \nonumber 
\een 
\subsection{Decoherence}
\label{ap:highd}
We begin our considerations with the decoherence factor. The mean is given by an expression
\ben
&&\left\langle \left\langle f^{\Gamma}_{T}(t;\omega_k)  \right\rangle\right\rangle= \\&& \frac{2 M \bar{\gamma}_0}{\hbar \pi \tau_T \omega_L \omega_U } \int^{\omega_U}_{\omega_L} \frac{1}{(\omega_k^2 - \Omega^2)^2}\bigg( \left( 1 -  \cos^2 \Omega t \right) + \frac{\Omega^2}{\omega_k^2} \left( 1 + \cos^2 \Omega t  \right) \\ &&- 2 \cos \Omega t \cos \omega_k t - \frac{2 \Omega}{\omega_k} \sin \Omega t \sin \omega_k t \bigg)  = \nonumber \\ &&\frac{2 M \bar{\gamma}_0}{\hbar \pi \tau_T \omega_L \omega_U }\left( I_1 + I_2 -2 I_3 -2 I_4 \right).
\een
Computing integrals we obtain:
\ben
&& I_1 =\int^{\omega_U}_{\omega_L} d \omega \frac{\Omega^2}{\omega^2(\omega^2 - \Omega^2)^2}\left( 1-\cos^2(\Omega t)\right) = \\ \nonumber &&\frac{\left( 1-\cos^2(\Omega t)\right)}{\Omega^2}\bigg(\frac{\omega_U-\omega_L}{\omega_U \omega_L } - \frac{\omega_U}{2(\omega_U^2-\Omega^2)} + \frac{\omega_L}{2(\omega_L^2-\Omega^2)} + \frac{3}{4\Omega} \log \frac{(\omega_U+\Omega)(\omega_L-\Omega)}{(\omega_U-\Omega)(\omega_L+\Omega)}  \bigg) \\ \nonumber \\
&&I_2 = \int^{\omega_U}_{\omega_L} d \omega \frac{1}{(\omega^2 - \Omega^2)^2}\left( 1+\cos^2(\Omega t)\right) = \\ \nonumber && \frac{1}{4 \Omega^2} \left( 1+\cos^2(\Omega t)\right)\bigg( \frac{2 \omega_L}{\omega_L^2-\Omega^2} - \frac{2 \omega_U}{\omega_U^2-\Omega^2}  + \frac{1}{\Omega} \log \frac{(\omega_U+\Omega)(\omega_L-\Omega)}{(\omega_U-\Omega)(\omega_L+\Omega)} \bigg)\\ \nonumber \\
&&I_3 = \int^{\omega_U}_{\omega_L} d \omega \frac{1}{(\omega^2 - \Omega^2)^2} \cos \omega t \cos \Omega t  = 
 \\ \nonumber && \frac{1}{4\Omega^2}\cos \Omega t\bigg[\frac{2 \omega_L \cos \omega_Lt}{\omega_L^2 - \Omega^2} - \frac{2 \omega_U \cos \omega_Ut}{\omega_U^2 - \Omega^2} + \\ \nonumber &&t \cos (\Omega t) F_{\text{Si}}(+,-,+,-) +   t \sin (\Omega t) F_{\text{Ci}}(+,-,-,+) + \\&& \nonumber \frac{1}{\Omega}(\cos\left(\Omega t \right) F_{\text{Ci}}(+,-,-,+) + \sin\left(\Omega t \right) F_{\text{Si}}(-,+,-,+)  \bigg] \\ \nonumber \\
&&I_4 = \int^{\omega_U}_{\omega_L} d \omega \frac{\Omega}{\omega(\omega^2 - \Omega^2)^2} \sin \omega t \sin \Omega t = \\&& \nonumber \frac{1}{2 \Omega^3} \sin \Omega t\bigg[
2\left(\text{Si}\left(\omega_Ut \right)-\text{Si}\left(\omega_Lt \right)\right) - \\ \nonumber &&\cos(\Omega t) F_{\text{Si}}(-,+,-,+)   - \sin(\Omega t) F_{\text{Ci}} (-,+,+,-) \\ \nonumber && + \frac{\Omega}{2}\bigg(\frac{2 \Omega \sin \omega_Lt}{\omega_L^2 - \Omega^2} - \frac{2 \Omega \sin \omega_Ut}{\omega_U^2 - \Omega^2} + \\ \nonumber  &&t \cos(\Omega t) F_{\text{Ci}}(-,+,+,-) +t \sin(\Omega t) F_{\text{Si}}(+,-,+,-) \bigg) \bigg] 
\een
Now we turn our attention to short-time behavior of the mean. We expand the above expressions up to the second order in time. This is a good approximation for $t \ll \omega_U^{-1}$. As a result we obtain  
\ben
&&I_1 = 
\\ \nonumber && \bigg(\frac{\omega_U-\omega_L}{\omega_U \omega_L } - \frac{\omega_U}{2(\omega_U^2-\Omega^2)} + \frac{\omega_L}{2(\omega_L^2-\Omega^2)} + \frac{3}{4\Omega} \log \frac{(\omega_U+\Omega)(\omega_L-\Omega)}{(\omega_U-\Omega)(\omega_L+\Omega)}  \bigg) t^2 + O(t^4)  \\
&&I_2 =
\\ \nonumber &&\frac{1}{4 \Omega^2} \bigg( \frac{2 \omega_L}{\omega_L^2-\Omega^2} - \frac{2 \omega_U}{\omega_U^2-\Omega^2}  + \frac{1}{\Omega} \log \frac{(\omega_U+\Omega)(\omega_L-\Omega)}{(\omega_U-\Omega)(\omega_L+\Omega)} \bigg) (2-\Omega^2 t^2) + O(t^4)  \\
&&I_3 =
\\ \nonumber &&\frac{\Omega^2}{4} \bigg( \frac{2 \omega_L}{\omega_L^2-\Omega^2} - \frac{2 \omega_U}{\omega_U^2-\Omega^2}  + \frac{1}{\Omega} \log \frac{(\omega_U+\Omega)(\omega_L-\Omega)}{(\omega_U-\Omega)(\omega_L+\Omega)}  \bigg) \\ \nonumber &&+ \frac{1}{2(\omega_L^2-\Omega^2)(\omega_U^2-\Omega^2)}\left( \omega_L \Omega^2 + \omega_U \omega_L - \omega_L \omega_U^2 -\omega_U \Omega^2 \right) t^2 + O(t^4)  \\
&&I_4 = \bigg( \frac{1}{2\omega_L^2-\Omega^2} - \frac{1}{2\omega_U^2-\Omega^2} + \frac{1}{4 \Omega} \log \frac{(\omega_U+\Omega)(\omega_L-\Omega)}{(\omega_U-\Omega)(\omega_L+\Omega)}   \bigg) t^2 + O(t^4) 
\een
Finally, we can add the terms to obtain
\ben
\label{ap:eq:htdst}
\left\langle \left\langle f^{\Gamma}_{T}(t_S;\omega)   \right\rangle\right\rangle= \frac{2 M \bar{\gamma}_0}{\hbar \pi \tau_T \omega_L \omega_U }  t^2 + O(t^4),
\een
what leads to the eq. (\ref{eq:dsh}) of the main text.

In the case of long-time behavior, one reaches a similar qualitative conclusion as for low-temperature case. Namely, for $t\gg (\omega_L - \Omega)^{-1}$ $I_3 \approx 0$ and $I_4 \approx 0$ so the only relevant terms are $I_1$ and $I_2$, what results in the following expression for long-time behavior of the mean    
\ben
\label{ap:htltd}
\left\langle \left\langle f^{\Gamma}_{T}(t_{AS};\omega) (\omega)   \right\rangle\right\rangle = \frac{2 M \bar{\gamma}_0}{\hbar \pi \tau_T \Delta \omega}\left(  A_{\Gamma} \cos^2 \left( \Omega t \right) + B_{\Gamma} \right) + O(t^{-1})
\een
where this time
\ben
&&A_{\Gamma} \equiv -\frac{1}{4\Omega^2}\bigg(\frac{\omega_U-\omega_L}{\omega_U \omega_L } + \frac{1}{2\Omega} \log \frac{(\omega_U+\Omega)(\omega_L-\Omega)}{(\omega_U-\Omega)(\omega_L+\Omega)}  \bigg), \nonumber \\
&&B_{\Gamma} \equiv \frac{1}{4\Omega^2}\bigg( \frac{\omega_L}{(\omega_L^2-\Omega^2)} - \frac{\omega_U}{(\omega_U^2-\Omega^2)} \bigg) - A_{\Gamma},\nonumber
\een
To obtain eq. (\ref{eq:htltdn}) of the main text one performs straightforward minimization of (\ref{ap:htltd}).

\subsection{Generalized overlap}
\label{ap:higho}
In the case of generalized overlap the mean is given by
\ben
\label{eq:ap:oht}
&&\left\langle \left\langle f^{B}_{T}(t;\omega_k)  \right\rangle\right\rangle = \\&& \nonumber \frac{2 M \bar{\gamma}_0 \tau_T}{ \hbar \pi \Delta \omega}  \int^{\omega_U}_{\omega_L} \frac{\omega^2}{(\omega_k^2 - \Omega^2)^2}\bigg( \left( 1 + \cos^2 \Omega t + \right) + \frac{\Omega^2}{\omega_k^2} \left( 1 -  \cos^2 \Omega t  \right) \\  &&- 2 \cos \Omega t \cos \omega_k t - \frac{2 \Omega}{\omega_k} \sin \Omega t \sin \omega_k t \bigg)  = \nonumber \\ &&\frac{2 M \bar{\gamma}_0 \tau_T}{ \hbar \pi \Delta \omega}\left( I_1 + I_2 -2 I_3 -2 I_4 \right) \nonumber
\een
The results of integration are:
\ben
I_1 = &&\int^{\omega_U}_{\omega_L} d \omega \frac{\omega^2}{(\omega^2 - \Omega^2)^2}\left( 1 + \cos^2 \Omega t  \right) = \\ \nonumber &&
\frac{\omega_L}{2(\omega_L^2-\Omega^2)} - \frac{\omega_U}{2(\omega_U^2-\Omega^2)} + \frac{1}{4\Omega}\log\frac{(\omega_U-\Omega)(\omega_L+\Omega)}{(\omega_L-\Omega)(\omega_U+\Omega)} \\
I_2 = &&\int^{\omega_U}_{\omega_L} d \omega \frac{\Omega^2}{(\omega^2 - \Omega^2)^2}\left( 1 - \cos^2 \Omega t  \right) = \\ \nonumber &&
\frac{\omega_L}{2(\omega_L^2-\Omega^2)} - \frac{\omega_U}{2(\omega_U^2-\Omega^2)} + \frac{1}{4\Omega}\log \frac{(\omega_U+\Omega)(\omega_L-\Omega)}{(\omega_L+\Omega)(\omega_U-\Omega)} \\
I_3=&&\int^{\omega_U}_{\omega_L} d \omega \frac{\omega^2}{(\omega^2 - \Omega^2)^2} \cos \omega t \cos \Omega t   = \\ \nonumber&&
 \frac{1}{4\Omega}\bigg(   \frac{2 \omega_L \cos \omega_Lt}{\omega_L^2 - \Omega^2} - \frac{2 \omega_U \cos \omega_Ut}{\omega_U^2 - \Omega^2} + \\&& \nonumber t (\cos (\Omega t)F_{\text{Si}}(+,-,+,-)  +  
 \sin (\Omega t) F_{\text{Ci}}(+,-,-,+))  +\\ \nonumber &&
 \frac{1}{\Omega} \com \left(\cos (\Omega t) F_{\text{Ci}}(-,+,+,-) + \sin (\Omega t) F_{\text{Si}}(+,-,+,-) \right) \bigg) \\
I_4 = &&\int^{\omega_U}_{\omega_L} d \omega \frac{\omega \Omega}{(\omega^2 - \Omega^2)^2} \sin \omega t \som = \\ \nonumber&&
 \frac{1}{4} \som \bigg[\frac{2 \Omega\sin \omega_L t} {\omega_L^2 - \Omega^2} - \frac{2 \Omega\sin \omega_U t} {\omega_U^2 - \Omega^2} + \\ \nonumber &&
t (\cos (\Omega t) F_{\text{Ci}}(-,+,+,-) + \sin (\Omega t) F_{\text{Si}}(+,-,+,-)) \bigg]
\een
With regard to to short-time behavior of the mean, we expand the above expressions up to the second order in time. This is a good approximation for $t \ll \omega_U^{-1}$. As a result we obtain
\ben
&&I_1 =  \\ &&  \left( \ul - \uu - \frac{1}{2\Omega} \loga \right)\frac{2-\Omega^2t^2}{2} + O(t^4) \nonumber \\
&&I_2 =  \\ &&  \frac{\Omega^2}{2}\left( \ul - \uu + \frac{1}{2\Omega} \loga \right) t^2 + O(t^4) \nonumber \\
&&I_3 =  \\ \nonumber && \frac{1}{2} \left( \ul - \uu - \frac{1}{2\Omega} \loga \right) +\\ \nonumber &&\frac{1}{4}\left( \uu (\omega_U^2+\Omega^2) - \ul (\omega_L^2+\Omega^2) + \right. \\ \nonumber &&\left.  3 (\omega_L-\omega_U)+2 \Omega \loga \right)t^2 + O(t^4)  \\
&&I_4 =  \\ \nonumber && \frac{\Omega^2}{2}\left( \ul - \uu - \frac{1}{2\Omega} \loga \right)t^2 + O(t^4) 
\een
Finally, we can add the terms to obtain
\ben
\label{ap:eq:htost}
\left\langle \left\langle f^{B}_{T}(t_S;\omega_k)  \right\rangle\right\rangle = && \ \frac{2 M \bar{\gamma}_0 \tau_T}{ \hbar \pi \Delta \omega}t^2 + O(t^4),
\een
what leads to the eq. (\ref{eq:osh}) of the main text.

After taking into account that for $t\gg (\omega_L - \Omega)^{-1}$ $I_3 \approx 0$ and $I_4 \approx 0$ so the only relevant terms are $I_1$ and $I_2$, one obtains the asymptotic formula for the mean 
\ben
\left\langle \left\langle f^{B}_{T}(t;\omega)   \right\rangle\right\rangle = &&\frac{2 M \bar{\gamma}_0 \tau_T}{ \hbar \pi \Delta \omega}\left(A_{B} \cos^2 \left( \Omega t \right) + B_{B} \right) + O(t^{-1}),
\een
with
\ben
&&A_{B} \equiv \frac{1}{2\Omega}\log\frac{(\omega_U-\Omega)(\omega_L+\Omega)}{(\omega_L-\Omega)(\omega_U+\Omega)} \\
&&B_{B} \equiv \frac{\omega_L}{\omega_L^2-\Omega^2} - \frac{\omega_U}{\omega_U^2-\Omega^2}
\een
After straightforward minimization of (\ref{eq:ap:oht}) one arrives at eq. (\ref{eq:htlton}) of the main text.

\end{document}